\documentclass[aps,prl,twocolumn,floatfix, showpacs]{revtex4}
\usepackage{graphicx}
\usepackage{dcolumn}
\usepackage{bm}
\begin{document}

\title{Reorientation of the large-scale circulation in turbulent Rayleigh-B{\'e}nard convection}
\author{Eric Brown}
\author{Alexei Nikolaenko}
\author{Guenter Ahlers}
\affiliation{Department of Physics and iQUEST, University of California, Santa Barbara, CA 93106}
\date{\today}
 
\begin{abstract}
We present measurements of the orientation $\theta_0(t)$ of the large-scale circulation (LSC) of turbulent Rayleigh-B{\'e}nard convection in cylindrical cells of aspect ratio 1.   $\theta_0(t)$ undergoes irregular reorientations. It contains reorientation events by rotation through angles $\Delta \theta$ with a monotonically decreasing  probability distribution $p(\Delta \theta)$, and by cessations (where the LSC stops temporarily) with   a uniform $p(\Delta\theta)$. Reorientations have Poissonian statistics in time.  The amplitude of the LSC and the magnitude of the azimuthal rotation rate have a negative correlation.

\end{abstract}


\maketitle

A major component of turbulent Rayleigh-B{\'e}nard convection \cite{Si94, Ka01, ahlers02} (RBC) in a fluid heated from below is the emission of hot (cold) volumes of  fluid known as ``plumes" from  bottom (top) thermal boundary layers. These plumes are carried by, and by virtue of their buoyancy in turn drive, a large-scale circulation (LSC) \cite{KH81, SWL89, CGHKLTWZZ89, CCL96,QT01a,TMMS05} known as the ``mean wind".  For cylindrical samples of aspect ratio $\Gamma \equiv D/L \simeq 1$ ($D$ is the sample diameter and $L$ the height) the LSC is a circulation in a near-vertical plane with up-flow and down-flow on opposite sides close to the side wall.  An interesting aspect of the LSC observed by others is that it can spontaneously and erratically change directions \cite{HYK91, CCS97, NSSD01, SBN02}, either by an azimuthal {\it rotation} of the entire structure without much change of the  flow speed,  or by reversing its flow direction without much rotation of the circulation plane. We shall refer to the latter process as {\it cessation} since it involves a momentary vanishing of the flow speed, and to both processes collectively as reorientations. Rotation through an angle $\Delta \theta = 1/2$ (we measure $\Delta \theta$ in units of $2\pi$, i.e. of revolutions) was observed in experiments \cite{CCS97}; but to our knowledge cessation so far had been demonstrated only in numerical simulations of a two-dimensional system \cite{HYK91} and in a dynamical-systems model of the phenomenon \cite{AGL05}. Thus, the re-orientation of the LSC dynamics is  still only partly understood.  Aside from its fundamental interest, the phenomenon is important for instance because it occurs in natural convection of the Earth's atmosphere \cite{DDSC00}, and because it is responsible for changes in the orientation of the Earth's magnetic field that is a result of convection reversal in the outer core \cite{GCHR99}.  

In this letter we describe experiments that shed new light on the nature of the reorientation of the LSC. We found that there is a broad spectrum of events leading to reorientation. At one extreme there is relatively slow rotation, over a time period of  order ten turnover times $\cal T$ of the LSC, without a significant decrease of the  circulation speed. At the other, there are near-instantaneous cessations that involve a vanishing of the flow speed followed by a re-initiation of the circulation with different angular orientations $\theta_0$. The re-orientations by either mechanism correspond to a continuous range of $\Delta \theta$, including the rotation by $\Delta \theta \simeq 1/2$ observed before \cite{CCS97} and the cessation involving $\Delta \theta \simeq 1/2$ that was seen in two-dimensional simulations \cite{HYK91} (where only this is possible), favored in the interpretation of some experiments \cite{SBN02}, and contained in a recently developed model \cite{AGL05}. As seen by Sreenivasan {\it et al.} \cite{SBN02}, the temporal succession of reorientations had a Poissonian distribution, suggesting that successive events are independent of each other.

The experiments cover the range $3\times10^8 \stackrel{<}{_\sim} R \stackrel {<}{_\sim} 10^{11}$ of the Rayleigh number $R \equiv \alpha g \Delta T L^3/\kappa \nu$ ($\alpha$ is the isobaric thermal expansion coefficient, $g$ the acceleration of gravity, $\Delta T$ the applied temperature difference, $\kappa$ the thermal diffusivity, and $\nu$ the kinematic viscosity). The Prandtl number was $\sigma \equiv \nu / \kappa = 4.4$.  Two cylindrical samples with aspect ratio $\Gamma  \simeq 1$ were used and are the medium and large sample described in detail elsewhere \cite{BNFA05}.  Both had circular copper top and bottom plates and a plexiglas side wall.  The medium (large) sample had  $D=24.81$ (49.67) cm and $L=24.76$ (50.61) cm.  Each was filled with water at  a mean temperature $T_m=40.0^{\circ} C$. Eight blind holes, equally spaced azimuthally in the horizontal mid-plane of the samples,  were drilled from the outside into the side walls. Thermistors were placed into them so as to be within $0.07$ cm of the fluid surface. They detected the location of the upflow (downflow)  of the LSC by indicating a relatively high (low) temperature, but did not perturb the flow structure in the fluid.  We made measurements with a sampling period $\delta t \simeq 6$ seconds, and fit the function $T_i = T_0 + \delta\cos(i \pi/4 - \theta_0), i = 0, \ldots,7$, separately at each time step, to the eight temperature readings. Here $\delta$ is a measure of the amplitude of the LSC and $\theta_0$ is the angular orientation of the plane of circulation. Typically the uncertainties were about 13\% for $\delta$ and 0.02 for $\theta_0$.   

\begin{figure}
\includegraphics[width=2.75in]{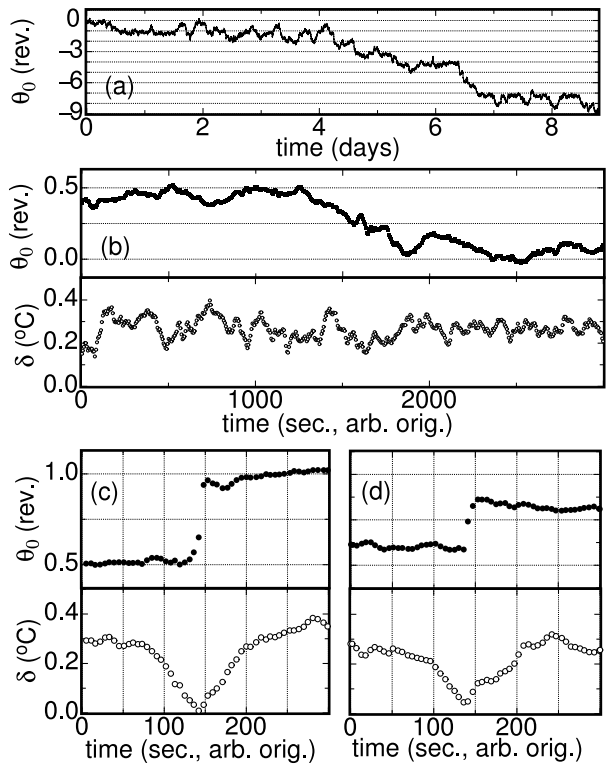}
\vskip 0.2in
\caption{Time series of the orientation $\theta_0$ (in revolutions) and of the amplitude $\delta$ (in $^\circ$C). (a): long time series of $\theta_0$ from the large sample for $R = 9.4\times 10^{10}$. (b) to (d): time series  during reorientations from the medium sample for $R = 1.1\cdot10^{10}$ for (b): a rotation; (c): a reversal; and (d): a half-reversal.}
\label{fig:reorientation_examples}
\end{figure}

In Fig. \ref{fig:reorientation_examples}a we show a long time series of $\theta_0$ from the large cell and for $R \simeq 10^{11}$. It covers a period of nine days. In this example, on average one sees net rotation, at a rate of about one revolution per day in the clockwise direction when viewed from above. Although $\theta_0(t)$ samples all values of $\theta$, there is an orientation $\theta_m$ where the probability distribution $p(\theta_0)$ has a maximum. Net rotation was observed also by Sun et al. \cite{SXX05}, but in the counter-clockwise direction. 

Hidden away in time series like that in Fig. \ref{fig:reorientation_examples} (a) are numerous reorientations. A few examples for the medium cell and $R = 1.1\cdot 10^{10}$ are shown in Figs. \ref{fig:reorientation_examples}b to d (for this case ${\cal T} =  49$s).  Figure \ref{fig:reorientation_examples} (b) shows a net angular change $\Delta \theta \simeq 1/2$ over a period of about $10{\cal T}$, similar to an event reported in Ref. \cite{CCS97}.  The reversal time much larger than ${\cal T}$ indicates a relatively slow rotation of the LSC orientation.   Throughout this event $\delta$ is seen to remain non-zero.

Figure \ref{fig:reorientation_examples} (c) also shows an angular change of approximately half a revolution, but the change occurred in a fraction of $\cal T$. Here $\delta$ decreased for about one turnover time before the reversal, reached approximately zero during the reversal, and then increased for about another $\cal T$.  One sees that the mean wind gradually slowed to a stop, reversed direction, and gradually sped up again without significant rotation of the plane of circulation.  This cessation is the mechanism favored by Sreenivasan et. al. \cite{SBN02} in the interpretation of their data. Presumably it would prevail in  systems where the rotational invariance is broken, either by imperfections of a nominally cylindrical sample or by a deliberate other choice of the geometry such as a square horizontal cross section. As we shall see below, in our system such events are relatively rare.

In Fig. \ref{fig:reorientation_examples} (d) the plots are qualitatively similar to (c). As in (c), the vanishing of $\delta$ indicates that the LSC stopped (also a cessation), but when it re-started, it assumed an orientation that differed by $\Delta \theta \simeq 1/4$ from the original one. Many other values of $\Delta \theta$ were observed, indicating that there is a larger class of phenomena with a continuous distribution of $\Delta \theta$. 

In order to automatically identify members of the expected large spectrum of reorientation events, we required reorientations to satisfy two criteria. First, the magnitude of the net angular change in orientation $| \Delta\theta |$ over a set of successive data points for $\theta_0$ had to be greater than $\Delta\theta_{min}$. Second,  the magnitude of the net average azimuthal rotation rate $|\langle\dot\theta_0\rangle| = |\Delta \theta / \Delta t|$ over that set had to be greater than $\dot\theta_{min}$.   Here $\Delta t$ is the duration of the reorientation. Often there were multiple overlapping sets that satisfied these requirements. Then the set with the maximum local quality factor $Q_n = |\Delta\theta| / (\Delta t)^n$ was chosen as the reorientation.  For $0 < n < 1$, $Q_n$ represents a compromise between choosing the maximum angular change ($Q_0$) or the maximum rotation rate ($Q_1$).  Any adjacent points to the chosen set were also included if the rotation rate $\dot\theta_0 = \delta\theta_0 / \delta t$ [$\delta\theta_0 = \theta_0(t+\delta t) - \theta_0(t)$] for the additional range was greater than $\dot\theta_{min}$ and of the same sign as the reorientation.  Mostly we used the parameters $\Delta\theta_{min} = 0.125$, $\dot\theta_{min} = 0.1 /  {\cal T}$, and $n = 0.25$; but since all three are arbitrary, we did the analysis over the ranges  $1/64 \le \Delta\theta_{min} \le 0.25$, $0.0125  \le \dot\theta_{min} {\cal T} \le 0.2 $, and $0 \le n \le 1$ to confirm that the results are not sensitive to the choice. For $\Delta\theta_{min}$ and $\dot\theta_{min}$, the smallest values used were close to the noise level of  the data, and above the largest values  too few reorientations were counted to yield useful statistics.  The qualitative conclusions of the analysis did not change with different parameter values. However, some of the quantitiave values changed. This will be mentioned as it is appropriate.

\begin{figure}
\includegraphics[width=2.75in]{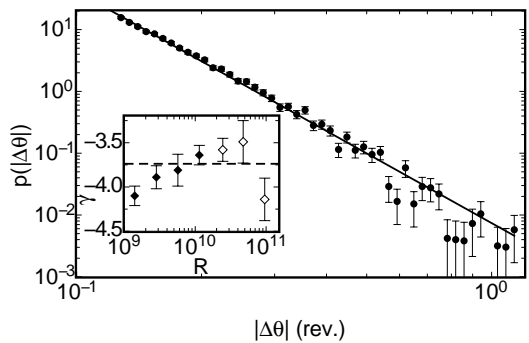}
\caption{Probability distribution of the angular change $p(\Delta\theta)$ for reorientations.  Solid circles: All data. Solid line: power-law fit to the data.  Inset: power-law exponent $\gamma$ versus $R$ for the medium cell  (solid diamonds) and large cell (open diamonds). Dashed line:  the value of $\gamma$ from the fit to all data.  Note that reversals with $\Delta \theta \simeq 0.5$ are relatively rare with $p(\Delta\theta = 1/2) \simeq 0.1$, implying that only about 1 \% of the events correspond e.g. to $\Delta\theta = 1/2 \pm 0.05$.}
\label{fig:prob_delta_theta}
\end{figure}

An obvious feature of reorientations is the angular change $\Delta\theta$.  All of the reorientations found for the medium sample were sorted into bins  according to $|\Delta\theta|$, regardless of the Rayleigh number $R$.  The probability distribution $p(|\Delta\theta|)$ is shown in Fig. \ref{fig:prob_delta_theta} with statistical error bars, in which the relative error is equal to the inverse square root of the number of reorientations in each bin.  Fitting all data for $p(|\Delta\theta|)$ by the maximum-likelihood mehod \cite{BR92}  to a power law in the range $|\Delta\theta| \ge 0.125$ yielded $p(|\Delta\theta|) \propto \left(|\Delta\theta|\right)^{\gamma}$, with $\gamma = -3.74\pm0.04$.  The same analysis was done also for reorientations at various fixed $R$.  These probability distributions also could be fitted by power laws, with the resulting $\gamma$ values shown for each $R$ in the inset of Fig. \ref{fig:prob_delta_theta}. One sees that the distribution was, within our resolution, independent of $R$.  The power-law exponents  were approximately constant over large ranges of the reorientation definition parameters; however they could vary by a factor of 2 for extreme values of the parameters. Qualitatively in all cases, the distribution was consistent with a power law for $|\Delta\theta| \ge 0.125$ with a large negative $R$-independent exponent.  The major qualitative conclusions are that there is a continuous distribution of $\Delta \theta$, that smaller reorientations are much more common than larger ones, and that there is no characteristic reorientation size.

\begin{figure}
\includegraphics[width=2.75in]{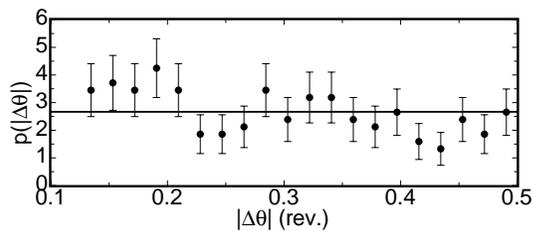}
\caption{Probability distribution of the angular change $p(|\Delta\theta|)$ for cessations. Here $|\Delta\theta|$ is reduced to the range $0 < |\Delta\theta| < 0.5$. Solid circles: All reorientations with $\langle\delta\rangle < \bar\delta(R) / 4$. Solid line: the uniform distribution.}
\label{fig:prob_delta_theta_small_amp}
\end{figure}

We would like to distinguish between cessations and rotations, so the probability distribution of the angular change $\Delta\theta$ for cessations only is plotted in Fig. \ref{fig:prob_delta_theta_small_amp}. There is not always a clear distinction between cessations and rotations, but here we used the criterion that a reorientation is a cessation when the average value $\langle\delta\rangle$ during the reorientation satisfied $\langle\delta\rangle < \bar\delta(R) / 4$, where $\bar\delta(R)$ is the average amplitude of an entire run at a given $R$.  This probability distribution for cessations is consistent with a uniform distribution, which is very different from the distribution for all reorientations shown in Fig. \ref{fig:prob_delta_theta}.  Since cessations make up only a small percentage of all reorientations, the probability distribution for all events as shown in Fig. \ref{fig:prob_delta_theta} still accurately describes rotations.  The uniform distribution of angular changes for cessations means that at the end of a cessation the orientation of the LSC is equally likely to be at any angle.  This suggests that the mean wind loses its memory of its initial orientation during the cessation.

\begin{figure}
\includegraphics[width=2.75in]{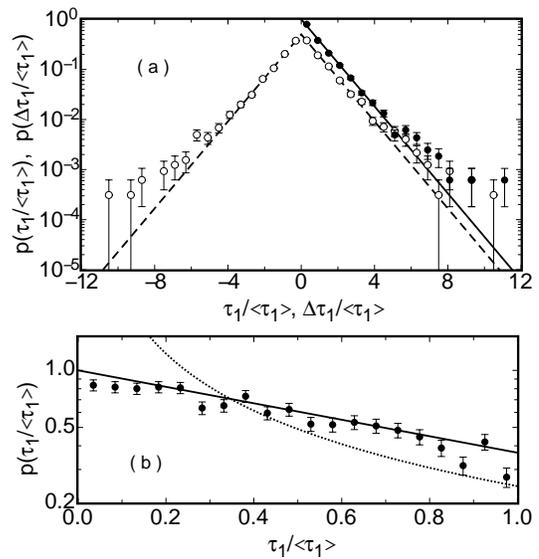}
\caption{Solid circles: probability distribution $p(\tau_1 / \langle\tau_1\rangle)$ of the time intervals $\tau_1$ between succesive reorientations, (a) over the entire range of $\tau_1$ measured, and (b) for $\tau_1 < \langle\tau_1\rangle$.  Open circles: the probability distribution $p(\Delta \tau_{1} / \langle\tau_1\rangle)$ of the difference between successive intervals.  Solid lines:  the function $p(\tau_1/\langle\tau_1\rangle) = \exp(-\tau_1/\langle\tau_1\rangle)$ representing the Poissonian distribution for $p(\tau_1/\langle\tau_1\rangle)$.  Dashed line:  the function $p(\Delta\tau_1/\langle\tau_1\rangle) = 0.5\exp(-\left|\Delta\tau_1/\langle\tau_1\rangle\right|)$ representing the Poissonian distribution for $p(\Delta\tau_1/\langle\tau_1\rangle)$.  Dotted line:  fit of a power law with exponent $-1$ to the data.}
\label{fig:prob_time_intervals}
\end{figure}

Another important aspect of reorientations is their distribution in time.  All of the time intervals $\tau_1$ between the $i$th and $(i+1)$th reorientations from the medium sample were sorted into bins according to the value of $\tau_1$.  The probability distribution  $p(\tau_1 / \left<\tau_1\right>)$ is  shown in Fig. \ref{fig:prob_time_intervals} (a), with statistical error bars as before.  The average time interval between reorientations $\left<\tau_1\right>$, computed separatly at each R,  was approximately  $30{\cal T}$, but quantitatively depended on the definition of reorientations and on $R$.  The data are in good agreement with the exponential function $p(\tau_1/\langle\tau_1\rangle) = \exp(-\tau_1/\langle\tau_1\rangle)$, which represents the Poissonian distribution.

The Poissonian nature of reorientations can be seen further by considering the difference between successive time intervals. Thus, $\Delta \tau_{1} = \tau_{1,i+1} - \tau_{1,i}$ was sorted into bins according to $\Delta \tau_{1}/\left<\tau_1\right>$ and the probability distribution $p(\Delta \tau_{1}/\left<\tau_1\right>)$ is also plotted in Fig. \ref{fig:prob_time_intervals} (a).  The data are in good agreement  with the function $p(\Delta\tau_1/\langle\tau_1\rangle) = 0.5\exp(-\left|\Delta\tau_1/\langle\tau_1\rangle\right|)$, which again represents the Poissonian distribution.  

The analysis in terms of $\tau_1$ and  $\Delta \tau_{1}$ implies that successive reorientation events are independent of each other.  For large $\tau_1/\langle\tau_1\rangle$ this is consistent with the conclusion of Sreenivasan et. al. \cite{SBN02} for reversals  in their experiment that were defined very differently.  However, these authors also found that for small time intervals, with $\tau_1 \stackrel{<}{_\sim} 30 {\cal T}$ in their experiment, that $p(\tau_1)$ followed a power law distribution with an exponent of $-1$.  To better compare the present work with this result, $p(\tau_1/\langle\tau_1\rangle)$ is plotted in the range $\tau_1 < \langle\tau_1\rangle \simeq 30 {\cal T}$ in Fig. \ref{fig:prob_time_intervals} (b).  Our data in this range are also consistent with the Poissonian distribution as found for the larger range, and inconsistent with the $-1$ power law found by Sreenivasan et. al.  This difference between the two experiments is likely due to the different methods of defining and analyzing events. We tested this by alternatively defining a reorientation as any crossing by $\theta_0(t)$ of an angle orthogonal to the preferred angle $\theta_m$ of the LSC orientation. This definition is more similar to that of Ref.~\cite{SBN02}. In that case we obtain a $p(\tau_{1}/\left<\tau_1\right>)$ that rises well above the Poisson distribution as $\tau_{1}/\left<\tau_1\right>$ decreases below one and that can be described quite well by a power law with an exponent close to one for $\tau_{1}/\left<\tau_1\right> < 1$.

\begin{figure}
\includegraphics[width=3.0in]{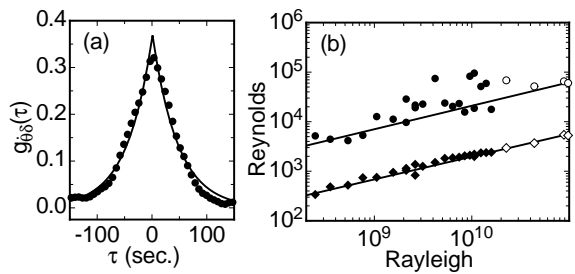}
\caption{(a): The negative cross-correlation $g_{\dot\theta\delta}(\tau)$ between the magnitude of the rotation rate $|\dot\theta|$ and the amplitude $\delta$ at $R = 1.1\cdot10^{10}$ (solid circles) and a fit by an exponential function (solid line).(b):  Reynolds numbers for the average $\tau_{g,0}$ [solid (open) circles] and width $\tau_{g,1}$ [solid (open) diamonds] of $g_{\dot\theta\delta}(\tau)$ for the medium(large) sample.}
\label{fig:corr_dthetadt_amp}
\end{figure}

Finally we consider the relationship between the magnitude of the azimuthal rotation rate $|\dot \theta_0|$ and amplitude $\delta$, since the examples in Fig. \ref{fig:reorientation_examples} show $|\dot\theta_0|$ is large when $\delta$ is small.   A negative cross-correlation function between the two quantities is given by
\begin{eqnarray} 
g_{\dot\theta\delta}(\tau)&=&-\langle(|\dot\theta_0(t)| - \langle|\dot\theta_0|\rangle)\left(\delta(t-\tau) - \bar\delta(R)\right)\rangle / (\sigma_{\dot\theta} \sigma_{\delta})\ ; \nonumber\\
\sigma_{\dot\theta}&=&\sqrt{\langle (|\dot\theta_0(t)| -\langle|\dot\theta_0|\rangle)^2 \ \rangle}\ ; \nonumber\\
\sigma_{\delta}&=&\sqrt{\langle (\delta(t) -\bar\delta(R))^2 \ \rangle}\ . \nonumber
\end{eqnarray}  
In Fig. \ref{fig:corr_dthetadt_amp} (a) we show $g_{\dot\theta\delta}(\tau)$ for $R = 1.1\cdot10^{10}$.  The function $g_{\dot\theta\delta}(\tau) = a_0\exp(-\left|(\tau - \tau_{g,0}) / \tau_{g,1}\right|)$ was fit to the data separately at each $R$ to yield $\tau_{g,0}(R)$ and  $\tau_{g,1}(R)$.  These time scales were made dimensionless by dividing the viscous diffusion time $L^2/\nu$ by each of them, yielding Reynolds numbers $R_e^{g,0} = L^2/(\tau_{g,0}\nu)$ and $R_e^{g,1} = L^2/(\tau_{g,1}\nu)$. These are plotted versus $R$ in Fig. \ref{fig:corr_dthetadt_amp} (b).  A power-law fit gives $R_e^{g,0} =0.41\cdot R^{0.47}$ and $R_e^{g,1} = 0.051\cdot R^{0.46}$.  One sees that these two power laws have approximately the same exponent. This exponent value is found also for the Reynolds number for $\cal T$ \cite{QT01b}, suggesting a relationship between the different time scales of the problem.  The time scale $\tau_{g,1}$ indicates how long the rotation rate and amplitude remain correlated, and it is about 80\% of ${\cal T}$.  The time scale $\tau_{g,0}$ indicates by how much the amplitude leads the rotation rate.  This is only about 6\% of ${\cal T}$, but the fact that the amplitude leads the rotation at all is remarkable, and suggests that an amplitude drop is a precursor of reorientations.

We are grateful to Detlef Lohse and Penger Tong for stimulating discussions. This work was supported by the US Department of Energy through Grant  DE-FG02-03ER46080.

\end{document}